# Isomerism in Quantum Dots: Geometries, Band Gaps, Dipole Moments, Ionization Energies and Heats of Formation

Prof. Vitaly V. Chaban

Federal University of São Paulo, São Paulo, Brazil

**Abstract**. Quantum dots (QDs) are applied in a variety of fields ranging from photovoltaics to biomedical imaging. Even the smallest QDs present a complicated potential energy surface characterized by a large set of stationary points. Each local minimum is an isomer of QD of given composition. An established theoretical methodology is hereby employed to obtain geometries of the QD isomers ($Cd_{16}Se_{16}$, $Cd_{16}Se_{16}$, $Zn_{16}S_{16}$, $Zn_{16}Se_{16}$) and predict their fundamental electronic and thermodynamic properties. Significantly scattered heats of formation, with an amplitude up to 1304 kcal mol$^{-1}$ in $Cd_{16}S_{16}$, were found for the most and the least thermodynamically stable isomers of QDs. The most shallow transition points can unlikely be observed in the experiments at finite temperature, since they are able to transform into more stable isomers upon thermal motion. Dipole moment is the most sensitive property to the QD isomer geometry. A global energy search technique was demonstrated to be an efficient tool to systematically identify isomers of QDs.

**Introduction**

Quantum dot (QD) is a common term to designate a semiconductor nanostructure that confines motion of its conduction band electrons and/or valence band holes in all directions. QDs are so small particles that their electronic and optical properties differ drastically from the bulk volume of the corresponding substance.[1-6] An entire QD can be seen as a single artificial atom because of its specific electronic structure (bound discrete electronic states). Many classical QDs are composed of the d-metal elements and the chalcogen elements. The radii of QDs do not exceed 5-6 nm, whereas the smallest QDs contain just a few dozens of atoms. Optoelectronic properties of QDs are rigorously predetermined by their size and elemental composition.

Existing and potential applications of QDs include solar cells, transistors, diode lasers, biomedical imaging, and even quantum computing. Thanks to their small radii, QDs can be suspended in a liquid solution relatively easily favoring less expensive and less time-consuming methods to prepare semiconductors for applications.[1,3,7-10]

The tunable absorption spectrum of QDs in conjunction with large extinction coefficients fosters their applications in light harvesting technologies. Some QDs are able to produce more than one exciton from one high-energy photon through carrier multiplication or multiple exciton generation.[11] Furthermore, the photovoltaics based on QDs should be cheaper to produce, since formation of QDs requires simple and adjustable chemical reactions. Aromatic self-assembled monolayers improve the band alignment at electrodes to enhance efficacy up to 10.7%.[12] Dipole moment and orientation of the self-assembled monolayer are important factors for the proposed band tuning. Techniques to precisely manipulate QDs and their derivatives at the molecular level are urgent to foster their applications in photovoltaics.

QDs exist in the form of a large number of geometrical isomers due to a variety of possible atom arrangement patterns. In principle, these isomers have different sets of thermodynamic and electronic properties, such as heat of formation, band gaps, dipole moments, atom-atom distances and angles, chemical reactivity, etc. It is by default assumed that the lowest-energy structure prevails in the experimental sample. This simplification does not, however, hold in all cases, especially when the solid-state structures are discussed. Method of preparation plays a paramount role in the determination of the most probable isomers. Since various QD structures correspond to the same elemental composition, understanding of all wide-spread isomers is important to accurately predict their physical chemical properties. The potential energy surface of any given elemental composition can be investigated theoretically by numerous individual geometry optimization runs and/or molecular dynamics simulations.

In the present work, a systematic isomer search is reported for the $Cd_{16}S_{16}$, $Cd_{16}Se_{16}$, $Zn_{16}S_{16}$, $Zn_{16}Se_{16}$ QDs. The resulting structures are characterized in terms of their fundamental properties, such as heat of formation ($H_f$), HOMO-LUMO band gap, dipole moment, and ionization energy. The differences and their magnitudes between the most thermodynamically stable isomers and the least thermodynamically stable isomers are highlighted.

**Methods and Methodology**

Electronic wave functions of the simulated particles were optimized at the semiempirical level of theory, Parametrized Model 7 (PM7).[13] The PM7 Hamiltonian is the newest semiempirical parametrization, which of based on the NDDO (Neglect of Diatomic Differential Overlap) approximation. PM7 can be seen as the Hartree-Fock method, in which certain integrals are substituted by the empirically known numbers. The 1- and 2-centered integrals are either evaluated approximately or parametrized if the corresponding reliable experimental data are available. In turn, the 3- and 4-centered integrals are ignored, since their role in most cases is not critical. The valence electrons are treated quantum mechanically, whereas an effect of the core electrons is reproduced by the computationally efficient pseudopotentials. Empirical corrections are also employed to more accurately simulate van der Waals interactions, peptide bonds, hydrogen bonds, etc. PM7 and earlier models were successfully applied to address versatile problems involving molecules, ions and crystals across the entire Periodic Table.[13-26] In this study, the self-consistence field convergence criterion was set to $10^{-5}$ Hartree.

While PM7 is known to predict accurate geometries and heats of formation, the electronic properties simulated by this method appear less accurate. For this reason, the reported electronic properties were rescaled with respect to the hybrid Becke-3-Lee-Yang-Parr density functional theory[27,28] in conjunction with the Dunning/Huzinaga valence double-zeta basis set[29] and Stuttgart pseudopotentials.[30] The reference values were obtained the corresponding global-minimum structures of each investigated QD (Table 1).

The potential energy surface was navigated according to the basin hopping algorithm.[31] An effective temperature to overcome potential barriers was set to 3000 K. The largest movement of a single atom per algorithm step was set to 0.5 Å. Each local optimization cycle was considered finished when none of the forces acting on the atoms exceeded 10 kJ mol$^{-1}$ Å$^{-1}$ (Broyden–Fletcher–Goldfarb–Shanno method). A total of 200 basin hopping steps were used to sample the potential energy surface of each QD.

Table 1. Electronic properties obtained for the global-minimum configurations of QDs by the B3LYP and PM7 methods. The observed discrepancies in electronic properties are expected, hence the systematically applied correction is validated.

| QD | $\mu_{PM7}$, D | $\mu_{B3LYP}$, D | $\Delta E_{PM7}$, eV | $\Delta E_{B3LYP}$, eV | $E_{ion,PM7}$, eV | $E_{ion,B3LYP}$, eV |
|---|---|---|---|---|---|---|
| $Cd_{16}S_{16}$ | 4.02 | 8.37 | 5.17 | 2.56 | 7.97 | 6.24 |
| $Cd_{16}Se_{16}$ | 11.3 | 7.14 | 5.64 | 2.56 | 7.92 | 6.15 |
| $Zn_{16}S_{16}$ | 26.4 | 16.6 | 5.16 | 1.39 | 6.85 | 5.60 |
| $Zn_{16}Se_{16}$ | 12.1 | 8.27 | 4.96 | 2.84 | 7.59 | 5.87 |

**Results and Discussion**

Large molecules, as a rule, contain a number of stationary points. Finding an arbitrary stationary point is straightforward by using the so-called local optimization methods. An identity of this local minimum must be verified by conducting a subsequent analysis of vibrational frequencies. Presence of negative frequencies means that a transition point was located, instead of a local minimum. Finding a global minimum is much more challenging and cannot be performed analytically. Numerous starts from arbitrary independent structures (geometries) are required to obtain a set of local minima. The deepest found local minimum can be subsequently assigned to be the global minimum for the given chemical composition. It is the researcher's responsibility to perform an enough number of local optimizations to ensure that the assigned global minimum exhibits the lowest possible potential energy.

Figure 1 depicts $H_f$ of the $Cd_{16}S_{16}$, $Cd_{16}Se_{16}$, $Zn_{16}S_{16}$, $Zn_{16}Se_{16}$ QDs obtained by the PM7 Hamiltonian. The sulfur containing QDs exhibit systematically less favorable $H_f$, as compared to the selenium containing QDs. In turn, the d-metal atoms do not have such a significant influence on $H_f$. In $Cd_{16}S_{16}$, $Cd_{16}Se_{16}$ and $Zn_{16}Se_{16}$, the global-minimum configuration was obtained in a single iteration of the basin hopping algorithm. In $Zn_{16}S_{16}$, the global-minimum geometry, -1239 kJ mol$^{-1}$, was obtained twice. Accordingly, the global minimum is not the most probable configuration, which could be reached starting from a specifically perturbed geometry. The same conclusion applies to the most shallow local-minimum configuration, whose fraction in the QD sample is expected to be marginal. Geometries and physical chemical properties of QDs are significantly determined by the method of their synthesis and less significantly by the method of storage. The discussed stationary points correspond to the absence of thermal motion.

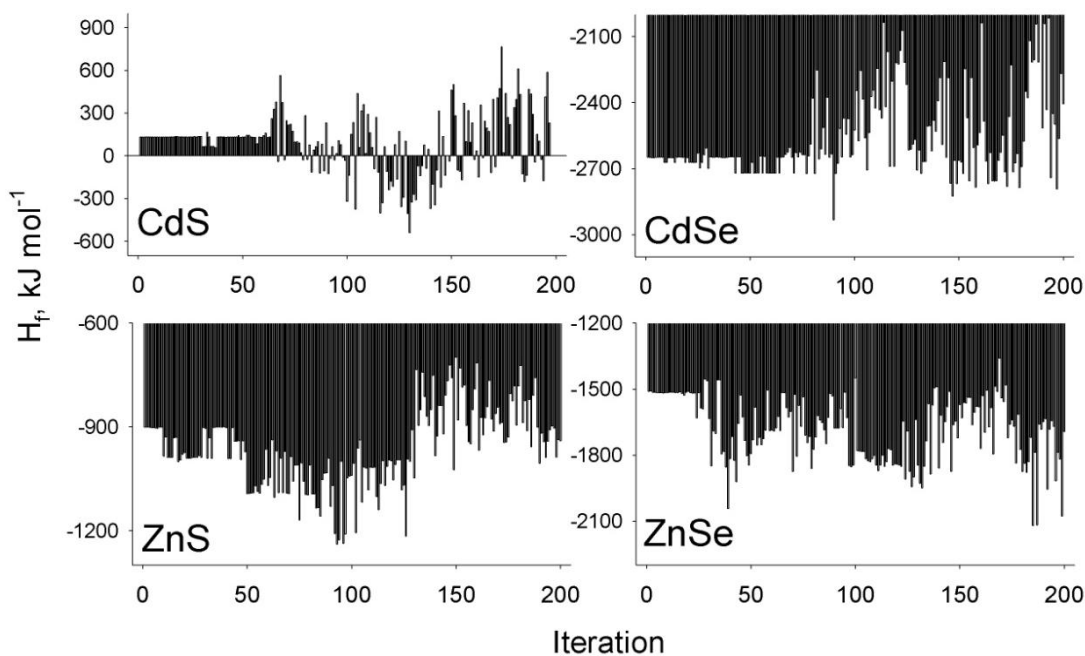

Figure 1. Standard heat of formation for different isomers of the investigated QDs.

All numerous unique $H_f$'s depicted in Figure 1 correspond to the stationary points on the potential energy surface. Therefore, these QD structures are chemically stable. The purpose of this work is to characterize a spectrum of possible QD structures and observe scattering of their fundamental properties. In terms of $H_f$, the difference between the most thermodynamically stable state (the smallest computed $H_f$) and the least thermodynamically stable state (the largest computed $H_f$) amounts to 1304 kJ mol$^{-1}$ ($Cd_{16}S_{16}$), 933 kJ mol$^{-1}$ ($Cd_{16}Se_{16}$), 540 kJ mol$^{-1}$ ($Zn_{16}S_{16}$), 760 kJ mol$^{-1}$ ($Zn_{16}Se_{16}$). Cadmium containing QDs, therefore, exhibit higher degree of scattering.

Band gaps for a large number of QD geometries and compositions are provided in Figure 2. The summarized band gaps were rescaled with respect to the B3LYP HDFT calculations of the global-minimum configuration, as described in the methodology. Significant scattering was found amounting to 2.0 eV ($Cd_{16}S_{16}$), 2.5 eV ($Cd_{16}Se_{16}$), 1.3 eV ($Zn_{16}S_{16}$), and 1.9 eV ($Zn_{16}Se_{16}$). PM7 provides systematically overestimated band gaps (Table 2) that must have been expected.

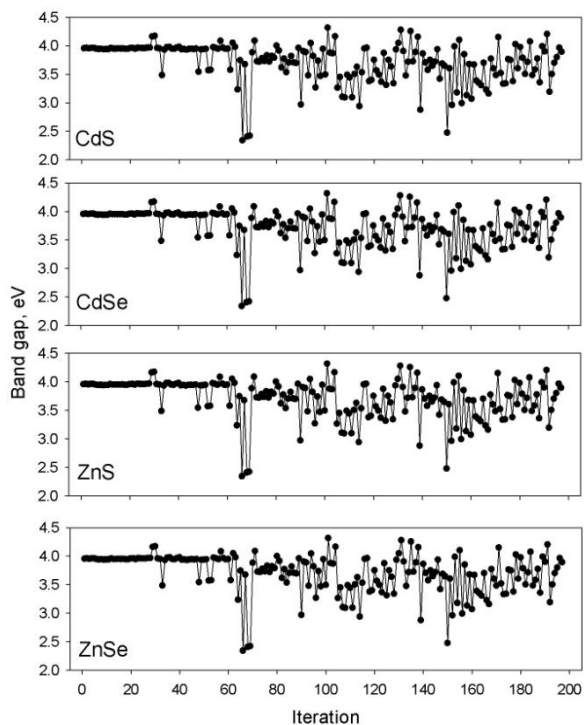

Figure 2. HOMO-LUMO band gap for different isomers of the simulated QDs.

The largest dipole moment was found to occur in one of the isomers of $Cd_{16}S_{16}$, $\mu = 37.7$ D (Figure 3). It corresponds to a highly polarized atomic arrangement, which appears nonetheless stable. In turn, the lowest dipole moment is just 0.39 D, which is found in one of the $Cd_{16}Se_{16}$ isomers. The global-minimum isomer exhibits an intermediate dipole moment $\mu = 7.1$ D. This heavy scattering suggests that it is very important to choose a correct isomer or a few probable isomers in the theoretical investigations of QDs. An arbitrary initial geometry of QD may provide a significant deviation.

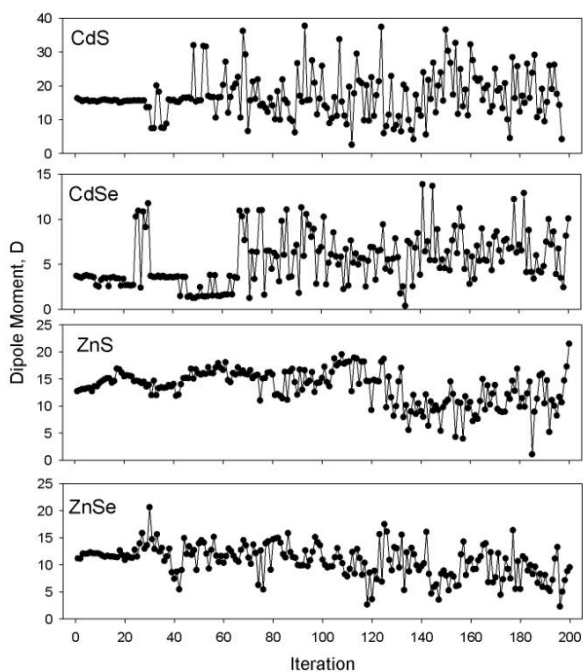

Figure 3. Dipole moment for different isomers of the simulated QDs.

The first ionization energy (Figure 4) characterizes an ability of a molecule to form a cation constituting an important descriptor of various isomers. The lowest ionization energy, 4.4 eV, was observed in the case of $Zn_{16}Se_{16}$, whereas the largest one, 6.6 eV, was observed in the case of $Cd_{16}S_{16}$. Remarkably, none of these boundary values the most or least thermodynamically favorable isomers in any composition. Thus, no rigorous correlation between thermodynamics and electronic properties was observed for most of the QD isomers.

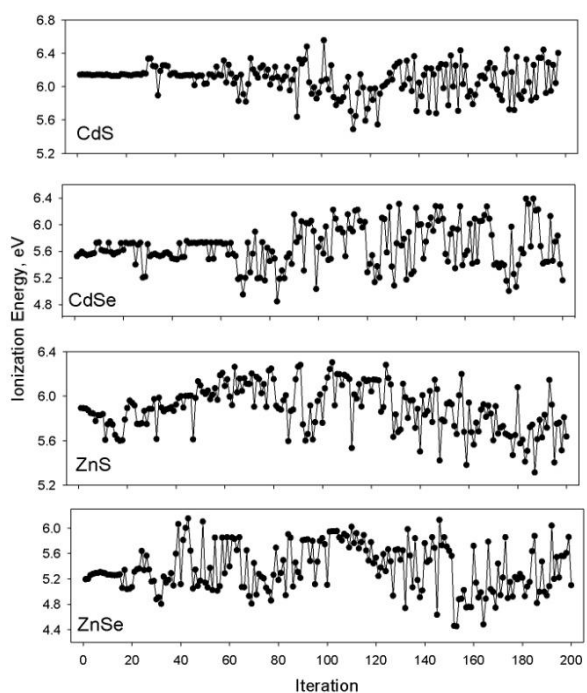

Figure 4. First ionization energy for different simulated QDs.

Figure 5 depicts the geometries of the global-minimum isomers. The unprotected QDs tend to adhere to spherical shapes, whereby metal-chalcogen bonds prevail in all QDs. Due to their small size, deviations from sphericity can be seen even in the most thermodynamically stable structures revealed by PM7. This is likely a natural phenomenon, while larger QDs (64, 128 atoms, etc) are more spherical.

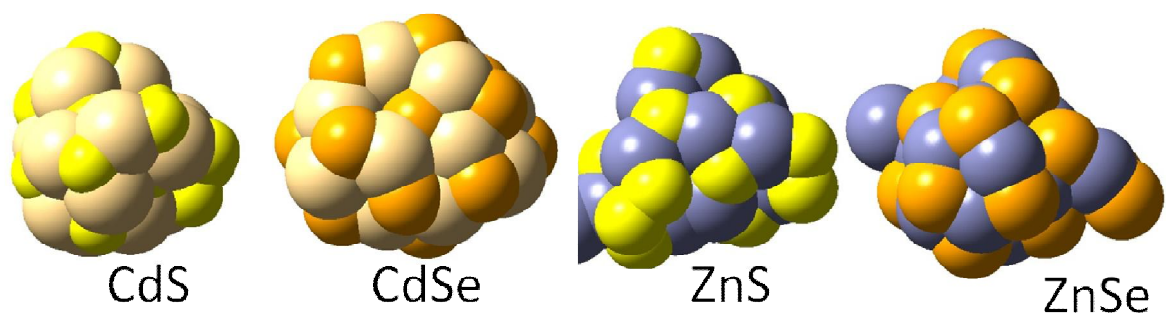

Figure 5. Global minimum geometries of the simulated QDs. Cadmium atoms are light pink; zink atoms are violet; sulfur atoms are yellow; selenium atoms are orange.

**Conclusions**

Potential energy surfaces of the $Cd_{16}S_{16}$, $Cd_{16}Se_{16}$, $Zn_{16}S_{16}$, $Zn_{16}Se_{16}$ QDs were extensively scanned to obtain up to 200 stationary point geometries for each QD. The optimized isomer featuring the lowest heat of formation was assigned to be the global minimum. Global-minimum geometries are most thermodynamically stable, but they are not necessary most abundant species, since geometries are significantly dependent on the synthesis of QDs. In turn, the isomers featuring the least negative heats of formation are expected to scarce in real samples, although they can maintain stability at low temperatures as local-minimum structures.

Fundamental thermodynamic and electronic properties were used to compare isomers with significantly different heats of formation. The most sensitive property is dipole moment, whose alteration from the least polar QD to the most QD reaches 1000%. In turn, the band gap alterations are much more modest, being nonetheless meaningful in a number of cases. For instance, the band gap of $Cd_{16}Se_{16}$ ranges 1.8 to 4.4 eV depending on which isomer is used for the calculation. Although many of the reported high-energy isomers are not stable at finite temperature, it is important to keep in mind when comparing different computational studies, which use different starting geometries and do not perform a comprehensive potential energy surface scanning to rate the isomers.

As any relatively large molecule, QDs exist in the form of many isomers, whose properties vary in the wide range. Stabilities of these isomers are determined by their thermodynamics and conditions of storage (temperature, pressure, chemical potential, etc). Global minimum search techniques are proven to be a valuable tool to identify isomers and compare their physical chemical properties.

## Author Information

E-mail for correspondence: vvchaban@gmail.com.